\documentstyle[aps,preprint,tighten,floats,epsf,feynmf]{revtex}

\newcommand{\tr}{\mbox{Tr}}

\begin{document}
\begin{fmffile}{pertxqcd_fmf}

\title{Perturbative Thermodynamics of Lattice QCD with
       Chiral-Invariant Four-Fermion Interactions}
\author{Michael Chavel\thanks{Electronic address: chavel@uiuc.edu}}
\address{
         Center for Theoretical Physics,
         Laboratory for Nuclear Science,
         and Department of Physics \\
         Massachusetts Institute of Technology, 
         Cambridge, MA 02139-4307\\
                  and\\
         Department of Physics\\
         University of Illinois,
         Urbana, IL 61801-3080}

\date{May 4, 1988}
\preprint{MIT-CTP-2737}
\maketitle

\begin{abstract}
Lattice QCD with additional chiral-invariant four-fermion interactions
is studied at nonzero temperature.  Staggered Kogut-Susskind quarks
are used.  The four-fermion interactions are implemented by
introducing bosonic auxiliary fields.  A mean field treatment of the
auxiliary fields is used to calculate the model's asymptotic
scale parameter and perturbative thermodynamics, including the
one-loop gluonic contributions to the energy, entropy, and pressure.
In this approach the calculations reduce to those of ordinary lattice
QCD with massive quarks.  Hence, the previous calculations of these
quantities in lattice QCD using massless quarks are generalized to the
massive case.
\end{abstract}

\pacs{12.38.Gc,12.38.Mh}

\section{Introduction}
\label{sec:intro}

One of the remaining challenges within the standard model of particle
physics is to understand the high temperature behavior of QCD with
light quarks.  This includes determining the location and order of the
chiral phase transition, associated with the restoration of the
$\mbox{SU}(N_f)_{\mbox{\scriptsize L}}\otimes
\mbox{SU}(N_f)_{\mbox{\scriptsize R}}$ flavor symmetry, and whether it
obeys the expected dimensional reduction
scenario\cite{Pisarski-Wilczek}.  In the high temperature plasma
phase, the status of the $\mbox{U}(1)_{\mbox{\scriptsize A}}$
symmetry, the equation of state, and the gluon screening masses,
necessary for high temperature perturbation theory, still remain to be
determined.

Lattice gauge theory appears to be an excellent nonperturbative tool
for studying these issues.  In fact, many aspects
of the deconfining phase transition and the high temperature plasma
phase in the gluonic sector are now well understood from Monte Carlo
studies of pure SU(3) gauge theory\cite{SU3}.  Unfortunately,
simulations of full QCD, including dynamical quarks, are much more
computationally intensive and so must be performed on smaller lattices
with relatively large lattice spacings and/or finite size effects.
Ref.~\cite{QCD_results} contains some recent reviews.

In an effort to more closely approach the true physics of continuum
QCD, many researchers have turned to using improved\cite{improved} or
``perfect''\cite{perfect} lattice actions, which reduce discretization
errors and lattice artifacts.  A somewhat different approach is the
$\chi$QCD model \cite{Brower,KogutLS,BarbourMK96,Chavel97}.  In this
model, extra auxiliary fields are introduced, which upon integration
yield chiral invariant four-fermion interactions of the form
${\textstyle{G\over2N_f/4}}[(\bar\psi\psi)^2 -
(\bar\psi\gamma_5\tau\psi)^2]$.  Such interactions are irrelevant to
the long distance physics and do not survive in the continuum limit.
However, when using the hybrid Monte Carlo algorithm the auxiliary
fields entering the fermion determinant allow one to avoid the usual
zero mass singularity when inverting the Dirac operator and perform
simulations directly in the chiral limit.  (i.~e.~the bare quark mass
can be set exactly to zero.)

To be specific, the $\chi$QCD action for $N_f$ flavors of staggered
quarks on an anisotropic lattice with spatial (temporal) lattice
spacing $a$ ($a_{\tau}$) is
\begin{eqnarray}
\label{xqcdaction}
S &=& \sum_x\biggl[\beta_{\sigma}\sum_{i<j}P_{ij}(x) +
      \beta_{\tau}\sum_jP_{0j}\biggr] +
      \sum_{a=1}^{N_f/4}\sum_{x,y}a^2a_{\tau}\bar\chi^a(x)
      Q(x,y)\chi^a(y) \nonumber\\
  &\ & \ {}+ {N_f\over8}\gamma\sum_{\tilde x}a^3a_{\tau}[\sigma^2(\tilde x) 
             + \pi^2(\tilde x)],
\end{eqnarray} 
with\footnote{
The model as considered here has a $\mbox{U}(1)_{\mbox{\scriptsize
L}}\otimes \mbox{U}(1)_{\mbox{\scriptsize R}}$ global symmetry,
generated by $1$ and $\gamma_5\tau_3$ ($\tau_3$ being the flavor
equivalent of $\gamma_5$).  One can also consider more physically
realistic $\mbox{SU}(N_f)_{\mbox{\scriptsize L}}\otimes
\mbox{SU}(N_f)_{\mbox{\scriptsize R}}$ symmetries.  However, for the
calculations considered in this paper, only the total number of quark
flavors is relevant.}
\begin{eqnarray}
Q(x,y) = \sum_{j=1}^3 {\cal M}_j(x,y) + \gamma_F{\cal
          M}_0(x,y) + \delta_{x,y}{1\over16}\sum_{<x,\tilde x>}a
          [\sigma(\tilde x)+i\varepsilon(x)\pi(\tilde x)]
\end{eqnarray} 
and $\gamma=1/G$. 
The staggered hopping term is
\begin{equation}
\label{hop}
{\cal M}_{\nu}(x,y)={1\over2} 
\eta_{\nu}(x)\bigl[U_{\nu}(x)\delta_{y,x+\hat\nu}
-U_{\nu}(x-\hat\nu)
\delta_{y,x-\hat\nu}\Bigr],
\end{equation}
where $\eta_\nu(x)\equiv
(-1)^{x_0+\cdots+x_{\nu-1}}$.  The auxiliary fields $\sigma(\tilde
x)$ and $\pi(\tilde x)$ are defined on the dual lattice sites.  The
symbol $<\! \tilde x,x \!>$ represents the $16$ dual sites $\tilde x$
adjacent to the direct lattice site $x$, and $\varepsilon(x)$ is the
alternating phase $(-1)^{x_0+x_1+x_2+x_3}$.  See \cite{auxillary} for
a more detailed discussion of the auxiliary field formulation used here.
$P_{\mu\nu}$ denotes the conventional Wilson plaquettes:
\begin{eqnarray}
\label{plaquette}
P_{\mu\nu} \equiv 1 - {1\over N}\mbox{Re}\mbox{Tr}
[U_{x,x+\mu}U_{x+\mu,x+\mu+\nu}U^{\dag}_{x+\nu,x+\mu+\nu}U^{\dag}_{x,x+\nu}].
\end{eqnarray}
The separate gauge couplings $\beta_{\sigma}=6/g^2_{\sigma}$,
$\beta_{\tau}=6/g^2_{\tau}$ as well as the extra parameter $\gamma_F$
are introduced in the usual way to maintain Euclidean invariance
in the continuum limit, when studying the model on anisotropic lattices
at nonzero temperature.

This paper studies the asymptotic scaling of the $\chi$QCD model and
its perturbative thermodynamics.  Section \ref{sec:scale} studies how
the four-fermion coupling affects the approach to the continuum limit
and the lattice scale parameter $\Lambda_{\mbox{\scriptsize L}}$.
Section \ref{sec:therm} deals with the perturbative thermodynamics of
the model and presents the one-loop gluonic corrections to the
energy, entropy, and pressure.  Throughout the paper, the four-fermion
interactions are dealt with via a mean field (or, equivalently,
$N_f\to\infty$) approximation, the details of which are given in
Sec.~\ref{sec:MFA}.  Treating the model in this way, the effects of
the four-fermion interactions on the asymptotic scaling behavior and
the perturbative thermodynamics are governed solely by the value of
the dynamical quark mass $m^2 = \langle\sigma\rangle^2 +
\langle\pi\rangle^2$.  Therefore, the one-loop gluonic corrections to
the model's thermodynamic observables reduce to those of standard
lattice QCD with a nonzero bare quark mass.  Since such calculations
have only appeared in the literature before at zero quark mass
\cite{Karsch82,Trinchero,Karsch89}, they are extended to nonzero
masses in Sec.~\ref{sec:pertQCD}.

\section{$\chi$QCD Scale Parameter}
\label{sec:scale}

Consider the $\chi$QCD model on an isotropic lattice ($a_{\tau}=a$), 
with $\beta_{\sigma}=\beta_{\tau}\equiv 6/g^2$ and $\gamma_F=1$.  In
the asymptotic scaling region of lattice QCD the gauge coupling $g$ is
related to the lattice spacing $a$ by
\begin{eqnarray}
\label{scaling}
a\Lambda_{\mbox{\scriptsize L}} = (\beta_0g^2)^{-\beta_1/2\beta_0^2}
\exp\biggl(-{1\over2\beta_0g^2}\biggr) \biggl[ 1 + O(g^2) \biggr],
\end{eqnarray}
which defines the conventional lattice QCD scale parameter
$\Lambda_{\mbox{\scriptsize L}}$.  In the continuum limit the scale
parameter is the only dimensional parameter. Hence, all physical
quantities measured on the lattice (such as particle masses, string
tension, etc.) must be expressible in terms of
$\Lambda_{\mbox{\scriptsize L}}$ by simple dimensional analysis.

The numerical value of $\Lambda_{\mbox{\scriptsize L}}$ depends upon
the specifics of the lattice regularization, and so is not universal.
The ratio of the lattice and continuum scale parameters, in any two
specific regularization schemes, is given by
\begin{eqnarray}
{\Lambda_{\mbox{\scriptsize L}}\over \Lambda_{\mbox{\scriptsize c}}} 
= Ma \exp\biggl[-{1\over2\beta_0}\biggl({1\over g^2_{\mbox{\scriptsize }}}
- {1\over g^2_{\mbox{\scriptsize c}} }\biggr)\biggr],
\end{eqnarray}
where $M$ is the continuum renormalization mass scale of interest
(momentum subtraction point or Pauli-Villars mass, etc.).  This ratio is
most easily calculated by a one-loop background field
calculation\cite{DashenGross}, where the gluonic action is expanded
around a slowly varying classical background configuration.

In $\chi$QCD one has the additional four-fermion coupling $G$, which
has physical dimensions of $(\mbox{length})^2$.  In the continuum
limit such an interaction is perturbatively irrelevant.  Therefore,
the asymptotic scaling formula (\ref{scaling}) must remain valid in
some window of $a\to 0$ for any given value of $G$.\footnote{
Unless, of course, their exists a second order phase transition at
some value of $G\equiv G_c$, where the four-fermion interaction
becomes both renormalizable and nontrivial.  In this case,
Eq.~(\ref{scaling}) may only hold when $G$ is taken smaller than
$G_c$.}
The numerical value of the scale parameter will change depending on
the particular value of $G$ used.  However, this dependence must
cancel out when any physical quantities are expressed in terms of the
continuum QCD scale parameter.  This will need to be verified in
future Monte Carlo simulations.

The effects of the explicit four-fermion interaction in $\chi$QCD are
most easily observed by a mean field calculation, or a large flavor
expansion, where the induced dynamical quark mass is related to the
mean values of the auxiliary fields  
(see Sec.~\ref{sec:MFA}).  The calculation of the lattice scale
parameter in lattice perturbation theory is then only affected by the
presence of this dynamical mass in the fermionic contributions to the
gluon self-energy (vacuum polarization).

The relevant diagrams are:
\begin{eqnarray}
\label{vac_a}
\parbox{30mm}{
\begin{center}
\begin{fmfgraph}(60,40)
\fmfpen{thin}
\fmfleft{i}
\fmfright{o}
\fmf{photon}{i,v1}
\fmf{fermion,tension=.5,left}{v1,v2,v1}
\fmf{photon}{v2,o}
\fmfdot{v1,v2}
\end{fmfgraph}
\end{center}
}
&=& {N_f\over4}{1\over4}\int_{q/2} 
      \mbox{Tr}[\gamma_{\mu}S_F(q-k/2)\gamma_{\nu}S_F(q+k/2)]
               \cos(q_{\mu}a_{\mu})\cos(q_{\nu}a_{\nu}),       
\\
\label{vac_b}
\parbox{30mm}{
\begin{center}
\begin{fmfgraph}(60,40)
\fmfpen{thin}
\fmfleft{i}
\fmfright{o}
\fmf{photon}{i,v}
\fmf{fermion}{v,v}
\fmf{photon}{v,o}
\fmfdot{v}
\end{fmfgraph}
\end{center}
} 
&=& -{N_f\over4}{1\over4}\int_{q/2} 
    \mbox{Tr}[\gamma_{\mu}S_F(q)]ia_{\mu}\sin(q_{\mu}a_{\mu})\delta_{\mu\nu}.
\end{eqnarray}
(The notation is explained in the appendix.)
Following \cite{Weisz80,Sharatchandra} and collecting together the
gauge, ghost, and fermion contributions, the ratio of the $\chi$QCD
scale parameter $\Lambda_{\chi}$ to the scale parameter
$\Lambda_{\mbox{\scriptsize MIN}}$ in the minimal subtraction scheme
can be written
\begin{eqnarray}
{\Lambda_{\chi}\over \Lambda_{\mbox{\scriptsize MIN}}} 
= \exp\biggl[J + {1\over\beta_0}\biggl({1\over48} 
               - 3P +{N_f\over2}P_5\biggr)\biggr],
\end{eqnarray}
with
\begin{eqnarray}
J &=& {\textstyle{1\over2}}(\ln4\pi - \gamma_{\mbox{\scriptsize Euler}}) 
= 0.9769042, \\
P &=& 0.0849780, \\
P_5 &=& \int_{-{\pi\over 2}}^{{\pi\over 2}}{d^4q\over (2\pi)^4}
{\cos^2(q_1)\cos^2(q_2)-{1\over3}\cos(2q_1)\cos(2q_2) \over
[\Delta_2(\xi) + m^2a^2]^2}  \nonumber \\
& \ & \qquad\qquad\qquad\qquad\qquad
  {}- {2\over3}\int_{-\infty}^{\infty}{d^4q\over (2\pi)^4}
\biggl[{1\over (q^2 +m^2a^2)^2} - {1\over (q^2 + 1)^2}\biggr].
\end{eqnarray}
Here, $m^2 = \langle\sigma\rangle^2 + \langle\pi\rangle^2$ is a
function of the two bare couplings $\{G,g\}$.  In simulations of
$\chi$QCD the value of the four-fermion coupling is to be chosen very
small, such that the observed values of $\langle\sigma\rangle$ and
$\langle\pi\rangle$ are principally due to {\it nonperturbative
effects of the gluons}.  Therefore, they cannot be reliably calculated
in perturbation theory and must be determined by Monte Carlo
simulation.  For this reason, it is best to keep $\Lambda_{\chi}$
parameterized by the dynamical quark mass
$ma=a\sqrt{\langle\sigma\rangle^2 + \langle\pi\rangle^2}$.

\begin{figure}
\begin{center}
\leavevmode
\epsfxsize=4.25in
\epsfbox{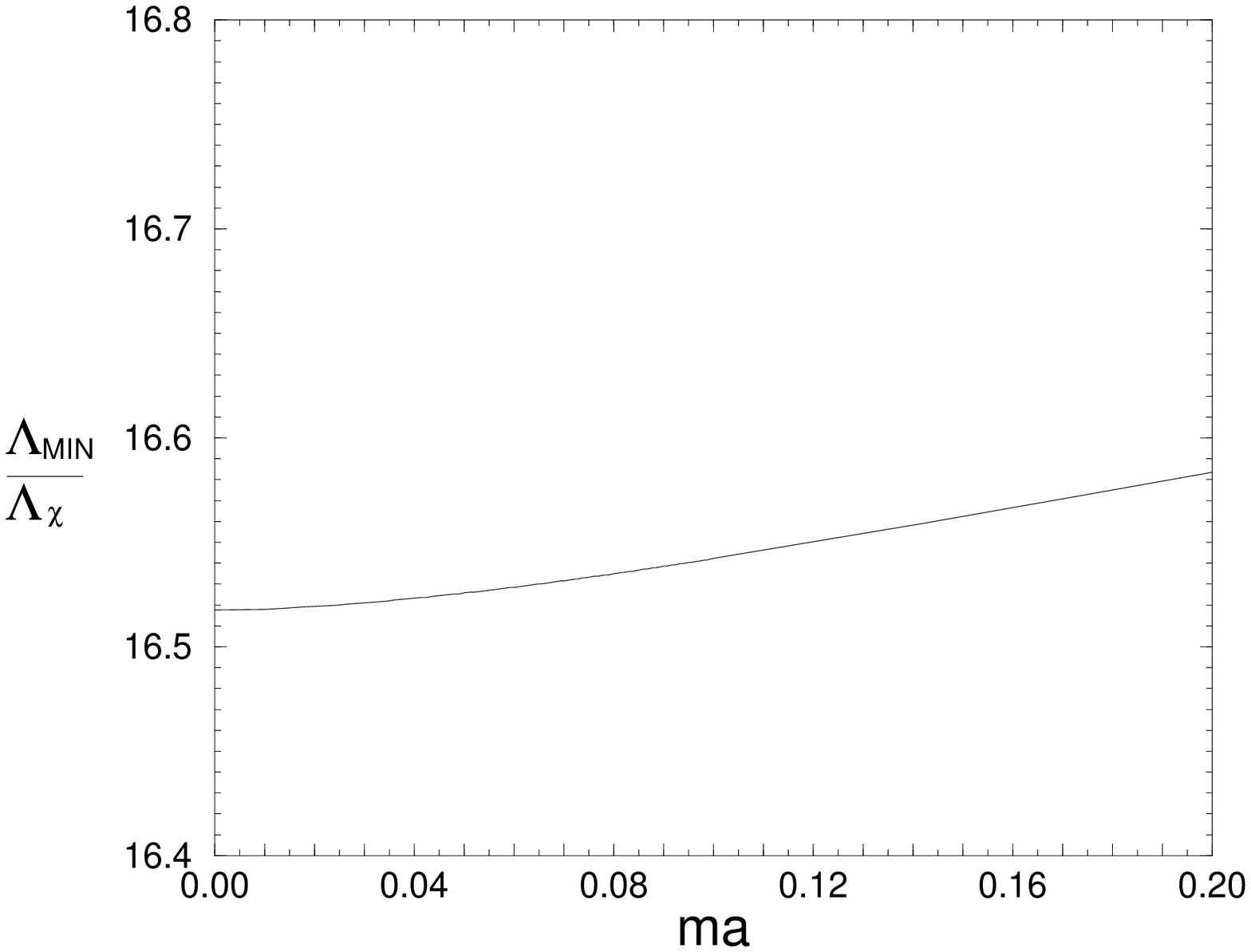}
\caption{${\Lambda_{\mbox{\scriptsize
MIN}}\over\Lambda_{\chi}}$ as a function of $ma$ with $N_F=2$.}
\label{fig:lambda_nf2}
\end{center}
\begin{center}
\leavevmode
\epsfxsize=4.25in
\epsfbox{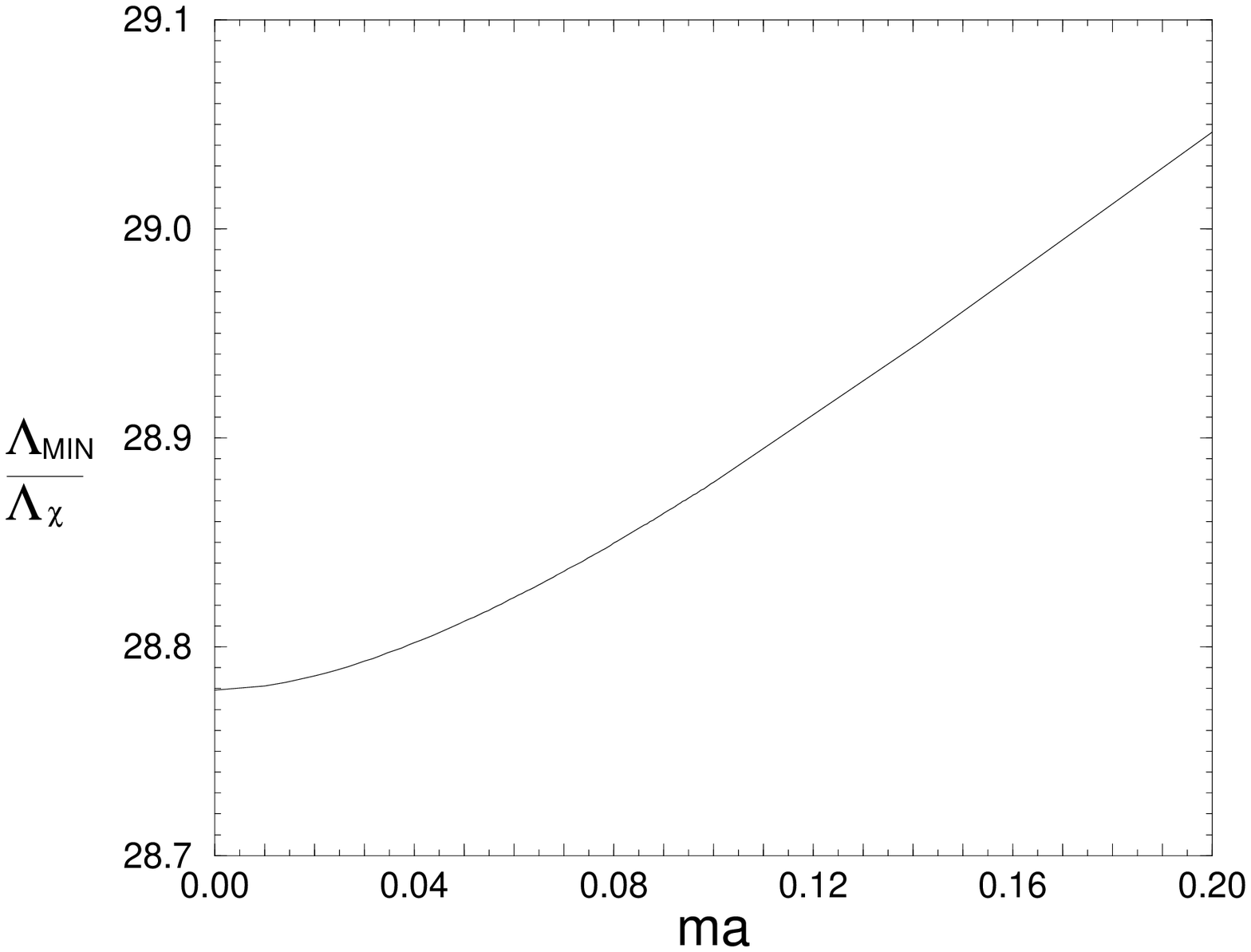}
\caption{${\Lambda_{\mbox{\scriptsize
MIN}}\over\Lambda_{\chi}}$ as a function of $ma$ with $N_F=4$.}
\label{fig:lambda_nf4}
\end{center}
\end{figure}

Values of $\Lambda_{\mbox{\scriptsize MIN}}/\Lambda_{\chi}$ are
plotted as a function of $ma$ in Figs.~\ref{fig:lambda_nf2} and
\ref{fig:lambda_nf4} for $N_f=2$ and $N_f=4$, respectively.
At $ma=0$ ($G=0$) I find
\begin{eqnarray}
{\Lambda_{\mbox{\scriptsize MIN}}\over\Lambda_{\chi}} =
{\Lambda_{\mbox{\scriptsize MIN}}\over\Lambda_{\mbox{\scriptsize L}}} =
\left\{
              \begin{array}{ll}
		10.846  & \ N_f=0  \\
                16.518  & \ N_f=2  \\
		28.779  & \ N_f=4				 
              \end{array}
        \right. ,
\end{eqnarray}
in agreement with \cite{Sharatchandra}.  As $ma$ increases the ratio
increases monotonically.  Assuming reasonable values such as
$\langle\psi\bar\psi\rangle\le 1$ and $\gamma\ge 10$, $ma$ will be 0.1
or less.  At $ma=0.1$ the increase in ${\Lambda_{\mbox{\scriptsize
MIN}}/\Lambda_{\chi}}$ is about 0.15\% (0.3\%) for 2 (4) fermion
flavors.  Thus, it is a relatively small effect which will be
difficult to extract from simulation data.

\section{$\chi$QCD Thermodynamics}
\label{sec:therm}

Working in the canonical ensemble, the Helmholtz free energy 
$F=-T\ln Z$, where $Z = \mbox{Tr} \exp{(-S/T)}$ is the lattice
regularized partition function. The temperature $T=
1/a_{\tau}N_{\tau}$, where $N_{\tau}$ ($N_{\sigma}$) is the number of
lattice sites in the temporal (spatial) direction(s).  Following the
usual convention define $\xi\equiv a/a_{\tau}$.  The energy density
and pressure are then given by
\begin{eqnarray}
\label{energydef}
\epsilon &=& -{1\over V} {\partial \ln Z \over \partial T^{-1}}
 = {T\over (aN_{\sigma})^3} \biggl\langle a_{\tau} 
    {\partial S\over\partial a_{\tau}}\biggr\rangle 
 = -T^4 \biggl({N_{\tau}\over\xi N_{\sigma}}\biggr)^3 
   \biggl\langle\xi{\partial S\over\partial\xi}\biggr\rangle, \\
p &=& T {\partial\ln Z\over\partial V} 
 = -{1\over3}{T\over (aN_{\sigma})^3} \biggl\langle 
   a{\partial S\over\partial a}\biggr\rangle    
 = -{T^4\over3} \biggl({N_{\tau}\over\xi N_{\sigma}}\biggr)^3 
   \biggl[\biggl\langle\xi{\partial S\over\partial\xi}\biggr\rangle 
    + \biggl\langle a{\partial S\over\partial a}\biggr\rangle\biggr]. 
\label{pressuredef}
\end{eqnarray}
Using the thermodynamic relations $F=U-TS$ and $p=-(\partial F/\partial
V)_T$ the entropy density can be expressed as
\begin{eqnarray}
\label{entropydef}
s = {\epsilon + p \over T} = {4\over3}{\epsilon\over T} -
 {T^3\over3} \biggl({N_{\tau}\over\xi N_{\sigma}}\biggr)^3
 \biggl\langle a{\partial S\over\partial a}\biggr\rangle.
\end{eqnarray}
The deviation from ideal gas behavior is typically characterized by 
\begin{eqnarray}
\delta \equiv \epsilon - 3p 
                    = T^4\biggl({N_{\tau}\over\xi N_{\sigma}}\biggr)^3 
 \biggl\langle a{\partial S\over\partial a}\biggr\rangle.
\end{eqnarray}

Before proceeding any further, note that Gaussian integration over the
auxiliary fields appearing in Eq.~(\ref{xqcdaction}) will yield a factor of
$(8\pi\xi/N_f\gamma a^4)^{N_{\sigma}N_{\tau}}$ in addition to the
desired four-fermion interactions.  Since the thermodynamic
observables are to be computed by differentiating the partition
function with respect to $a$ and $\xi$, it is best to remove this
factor from $Z$.  This can be accomplished by adding to the action the
extra term
\begin{equation}
\label{freeterm}
\gamma{N_f\over8}\sum_{\tilde x} a^3a_{\tau}
\bar\eta(\tilde x)\eta(\tilde x), 
\end{equation}
where $\bar\eta$ and $\eta$ free Grassmannian fields.  Integration over
$\bar\eta$ and $\eta$ will then cancel the cancel the corresponding
contribution from the auxiliary fields.  Of course, if one is only
concerned with measuring quantities relative to the zero temperature
vacuum this becomes unnecessary.
 
Adding (\ref{freeterm}) to the $\chi$QCD action
(\ref{xqcdaction}) it follows immediately that
\begin{eqnarray}
\label{dS/dxi}
\biggl\langle\xi{\partial S\over\partial\xi}\biggr\rangle &=&
N_{\sigma}^3N_{\tau}\biggl[
\xi{\partial\beta_{\sigma}\over\partial\xi}\langle P_{ss}\rangle +
\xi{\partial\beta_{\tau}\over\partial\xi}\langle P_{st}\rangle
+ \biggl(-1 + \xi{\partial Z_{\psi}\over\partial\xi}\biggr)
 {N_f\over4}\langle\bar\chi Q\chi\rangle
\nonumber \\ &\ & {}+
\xi{\partial\gamma_F\over\partial\xi}{N_f\over4}\langle\bar\chi{\cal
M}_0\chi\rangle + \biggl(-\gamma 
+\xi{\partial\gamma\over\partial\xi}\biggr){N_f\over8}
\biggl(\langle\sigma^2+\pi^2\rangle +
\langle\bar\eta\eta\rangle\biggr) 
\biggr].
\end{eqnarray} 
All the averages on the right hand side are per space-time volume.
$\langle P_{ss}\rangle$ ($\langle P_{st}\rangle$) stands for the
average value of the space-space (space-time) plaquettes.  Note that
$\langle\bar\chi Q\chi\rangle = -3$ and $\gamma
{\textstyle{N_f\over8}}\langle\bar\eta\eta\rangle= -1$.

In order to calculate the regularization dependence of the bare
parameters appearing above, a mean field treatment of the auxiliary
fields together with perturbative QCD corrections will be used.  The
details are presented in Secs.~\ref{sec:MFA} and \ref{sec:pertQCD}.
In this approach the calculations of
$\partial\beta_{\sigma}/\partial\xi$,
$\partial\beta_{\tau}/\partial\xi$, and $\partial\gamma_F/\partial\xi$
become the same as in ordinary lattice QCD with
$\langle\sigma\rangle^2+\langle\pi\rangle^2$ substituting for the
square of the bare quark mass.  The quantity $\partial
\gamma/\partial\xi$ is determined by absorbing the $\xi$
dependence of the one-loop gluonic corrections to the
$\sigma\bar\psi\psi$ vertex
into the value of $G=1/\gamma$.  This insures that the dynamical
fermion mass, given by the pole in the quark propagator, is invariant
with respect to $\xi$. 

In Sec.~\ref{sec:MFA} it is shown that
\begin{eqnarray}
\label{dgamma/dxi}
{1\over\gamma}\biggl(\xi{\partial\gamma\over\partial\xi}\biggr)_{\xi=1} =
  -2g^2\biggl({\partial C_m\over\partial\xi}\biggr)_{\xi=1}.
\end{eqnarray}
The quantity $(\partial C_m/\partial\xi)_{\xi=1}$ is plotted in
Fig.~\ref{fig:dcmdq} as a function of
$ma=a\sqrt{\langle\sigma\rangle^2+\langle\pi\rangle^2}$.
It is the only quantity of interest which varies significantly with
$ma$.  [($\partial C_m/\partial\xi)_{\xi=1}$ varies about 2\% from
$ma=0$ to 0.1, while $\partial C_{\sigma}/\partial\xi$, $\partial
C_{\tau}/\partial\xi$ and $\partial C_F/\partial\xi$ all vary 0.2\% or
less over the same range.]  Hence, the values of $\partial
\beta_{\sigma}/\partial\xi$, $\partial \beta_{\tau}/\partial\xi$ and
$\partial \gamma_F/\partial\xi$ calculated with massless quarks can
safely be substituted.

Differentiating the action with respect to the spatial lattice spacing
yields
\begin{eqnarray}
\label{dS/da}
\biggl\langle a{\partial S\over\partial a}\biggr\rangle &=&
N_{\sigma}^3N_{\tau}\biggl[
a{\partial\beta_{\sigma}\over\partial a}\langle P_{ss}\rangle +
a{\partial \beta_{\tau}\over\partial a}\langle P_{st}\rangle 
+  \biggl(3 + {\partial Z_{\psi}\over\partial\ln a}\biggr)
 {N_f\over4}\langle\bar\chi Q\chi\rangle
+ a{\partial \gamma_F\over\partial a}
          {N_f\over4}\langle\bar\chi{\cal M}_0\chi\rangle 
\nonumber \\
&\ & {}+ \biggl(4\gamma +a{\partial\gamma\over\partial a}\biggr)
{N_f\over8}\biggl(\langle\sigma^2 + \pi^2\rangle + 
\langle\bar\eta\eta\rangle\biggr)
+ {N_f\over4}\langle\bar\chi(\sigma+i\varepsilon\pi)\chi\rangle\biggr].
\end{eqnarray} 
Using the identity
${\textstyle{N_f\over4}}\langle\bar\chi(\sigma+i\varepsilon\pi)\chi\rangle
= 2(1-\gamma {\textstyle{N_f\over8}}\langle\sigma^2 + \pi^2\rangle)$
the last two terms can be combined.  To lowest order in $g$, the
values of ${\partial\beta_{\sigma}/\partial a}$, ${\partial
\beta_{\tau}/\partial a}$ and ${\partial \gamma_F/\partial a}$ are the
same as in ordinary lattice QCD, since they are controlled by the
lowest universal coefficient of the beta function (see
Sec.~\ref{sec:pertQCD}).  There is no bare quark mass in the action.
The expected scaling of such a  mass is taken over here by
the scaling of the four-fermion coupling.  
In Sec.~\ref{sec:MFA} it is shown that
\begin{eqnarray}
\label{dgamma/da}
{1\over\gamma}\biggl(a{\partial\gamma\over\partial a}\biggr)_{\xi=1} 
              = -g^2{2\over2\pi^2} + O(g^4). 
\end{eqnarray}

Measuring the energy density and pressure with respect to the zero
temperature vacuum, by subtracting measurements on a symmetric lattice
(denoted by $\langle\cdots\rangle_{sym}$), the final expressions at
$\xi=1$ are
\begin{eqnarray}
\epsilon &=& T^4 N_{\tau}^4 
  \biggl\{ {6\over g^2}\biggl(1 - g^2{\partial C_{\sigma}
  \over\partial\xi}\biggr)  \biggl[\langle P_{ss}\rangle
 - \langle P_{ss}\rangle_{sym}\biggr] 
  + {6\over g^2}\biggl(-1 - g^2{\partial C_{\tau}
  \over\partial\xi}\biggr)\biggl[\langle P_{st}\rangle 
  - \langle P_{st}\rangle_{sym}\biggr] 
  \nonumber \\ 
 & \ &\quad {}+ \biggl(1 + g^2{\partial
  C_F\over\partial\xi}\biggr){N_f\over4} \biggl[\langle\tr{\cal
  M}_0 Q^{-1}\rangle - \langle\tr{\cal M}_0 Q^{-1}\rangle_{sym}
  \biggr] \nonumber \\ 
 & \ & \quad {}+ \biggl(1 + 
  g^2 2{\partial C_m\over\partial\xi}\biggr)\gamma{N_f\over8}
  \biggl[\langle\sigma^2+\pi^2\rangle
  -\langle\sigma^2+\pi^2\rangle_{sym}\biggr] \biggr\},
  \label{xqcd-energy}\\ 
p &=& {\epsilon\over3} + {T^4\over3} N_{\tau}^4 \biggl\{
  12\beta_0\biggl[\langle P_{ss}\rangle-\langle P_{ss}\rangle_{sym} 
  + \langle P_{st}\rangle - \langle P_{st}\rangle_{sym}\biggr] 
  \nonumber \\
 & \ &\quad {}+
  \biggl(-2+g^2{2\over2\pi^2}\biggr)\gamma{N_f\over8}
  \biggl[\langle\sigma^2+\pi^2\rangle
  -\langle\sigma^2+\pi^2\rangle_{sym}\biggr] \biggr\}.
  \label{xqcd-pressure} 
\end{eqnarray}
The (absolute) entropy density at $\xi=1$ is
\begin{eqnarray}
\label{xqcd-entropy}
s &=& {4T^3\over3}  N_{\tau}^4 \biggl\{ 
{6\over g^2}\biggl[1 + g^2{1\over2}\biggl({\partial C_{\tau}
  \over\partial\xi}-{\partial C_{\sigma}\over\partial\xi}\biggr)
  \biggr] \biggl[\langle P_{ss}\rangle -\langle P_{st}\rangle\biggr] 
  \nonumber \\ 
 & \ &\quad {}+ \biggl(1 + g^2{\partial
  C_F\over\partial\xi}\biggr){N_f\over4} \biggl[\langle\tr{\cal
  M}_0 Q^{-1}\rangle 
 - {3\over4} \biggr] \nonumber \\ 
 & \ & \quad {}+
  {1\over2}\biggl(1+g^2{1\over2\pi^2}+g^2
  4{\partial C_m\over\partial\xi}\biggr)
  \biggl[\gamma{N_f\over8}\langle\sigma^2+\pi^2\rangle
 - 1 \biggr] \biggr\}. 
\end{eqnarray}
In writing the above expressions I have used the identity
$\langle\bar\chi A \chi\rangle = -\langle\tr A Q^{-1}\rangle$.
The relations (\ref{gauge_sum_rule}) and (\ref{self-energy_sum_rule})
from Sec.~\ref{sec:pertQCD} have also been used to simplify the
expression for the entropy density.  For
$ma=a\sqrt{\langle\sigma\rangle^2+\langle\pi\rangle^2}$ less than 0.1
the values of $C_{\sigma}$, $C_{\tau}$, and $C_F$ are effectively the
same as in ordinary QCD:
\begin{eqnarray}
\biggl({\partial C_{\sigma}\over\partial\xi}\biggr)_{\xi=1} 
&=& 0.20161 - {N_f\over2}0.00062, \\
\biggl({\partial C_{\tau}\over\partial\xi}\biggr)_{\xi=1}
&=&  -0.13195 - {N_f\over2}0.00782, \\
\biggl({\partial C_F\over\partial\xi}\biggr)_{\xi=1} &=& -0.2132.
\end{eqnarray}
They are discussed in full detail in Sec.~\ref{sec:pertQCD}.  Values
of $(\partial C_m \partial\xi)_{\xi=1}$ are tabulated in
Table~\ref{table:dcmdq} for values of $ma < 0.2$.  

Note that $\gamma{\textstyle{N_f\over8}}\langle\sigma^2+\pi^2\rangle$
approaches the Gaussian free field value of $1$ as $\gamma\to\infty$.
In which case, Eqs.~(\ref{xqcd-energy}) -- (\ref{xqcd-entropy}) yield
the correct expressions for massless QCD.  Note also that in this
model
$\langle\tr Q Q^{-1}\rangle = 3$ and
${\textstyle{N_f\over4}}\langle\tr(\sigma+i\varepsilon\pi)Q^{-1}\rangle
= 2(\gamma {\textstyle{N_f\over8}}\langle\sigma^2+\pi^2\rangle - 1)$.
Therefore, 
\begin{equation}
\label{sym_ward_identity}
\langle\tr{\cal M}_0 Q^{-1}\rangle_{sym} =
{\textstyle{3\over4}} - {\textstyle{2\over N_f}}
(\gamma{\textstyle{N_f\over8}}\langle\sigma^2+\pi^2\rangle_{sym}-1).
\end{equation}

As of this writing the only numerical study of $\chi$QCD is that of
Ref.\cite{KogutLS}, with $N_f=2$ massless flavors of quarks on a
$8^3\times 4$ lattice at $\gamma=10$. In that study $T=0$ measurements
(on a $8^3\times 24$ lattice) were only made at $6/g^2=5.4$.  At that
value of the gauge coupling they found $\langle\tr{\cal M}_0
Q^{-1}\rangle-{\textstyle{3\over4}}=-0.0200(3)$ and
$\gamma{N_f\over8}\langle\sigma^2+\pi^2\rangle=1.0198(2)$
\cite{KogutLS98-prvt}, in good agreement with
Eq.~(\ref{sym_ward_identity}) and the vanishing of the entropy at zero
temperature.  Further zero temperature measurements would be necessary
to extract the energy density and pressure.

\begin{figure}
\begin{center}
\leavevmode
\epsfxsize=4.2in
\epsfbox{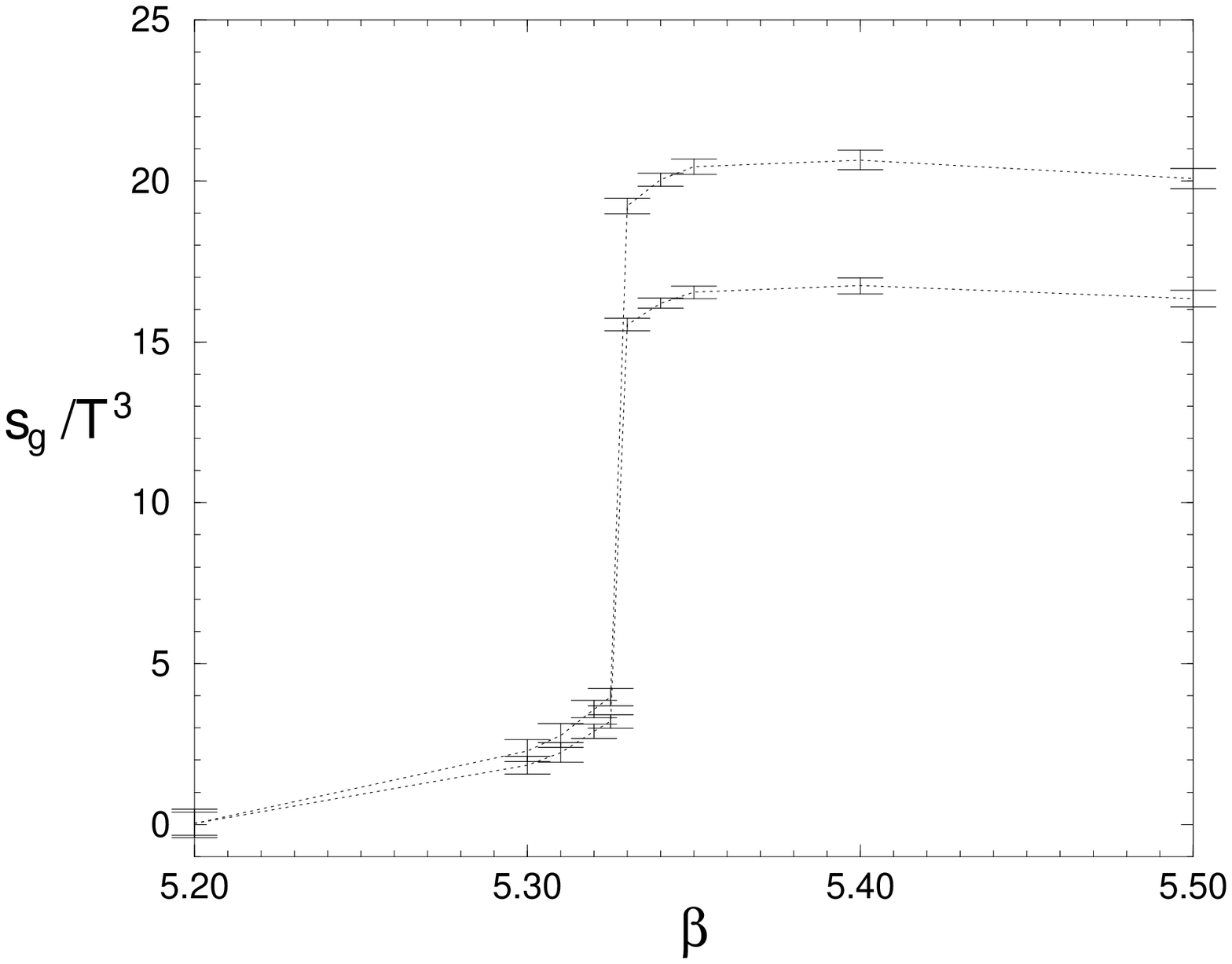}
\caption{Gluonic entropy density using the data from \protect\cite{KogutLS},
         for two massless flavors on a $8^3\times4$ lattice at
         $\gamma=10$.  The upper line is for the tree level results and the
         lower line the one-loop corrected results.}
\label{fig:glue_entropy}
\leavevmode
\epsfxsize=4.2in
\epsfbox{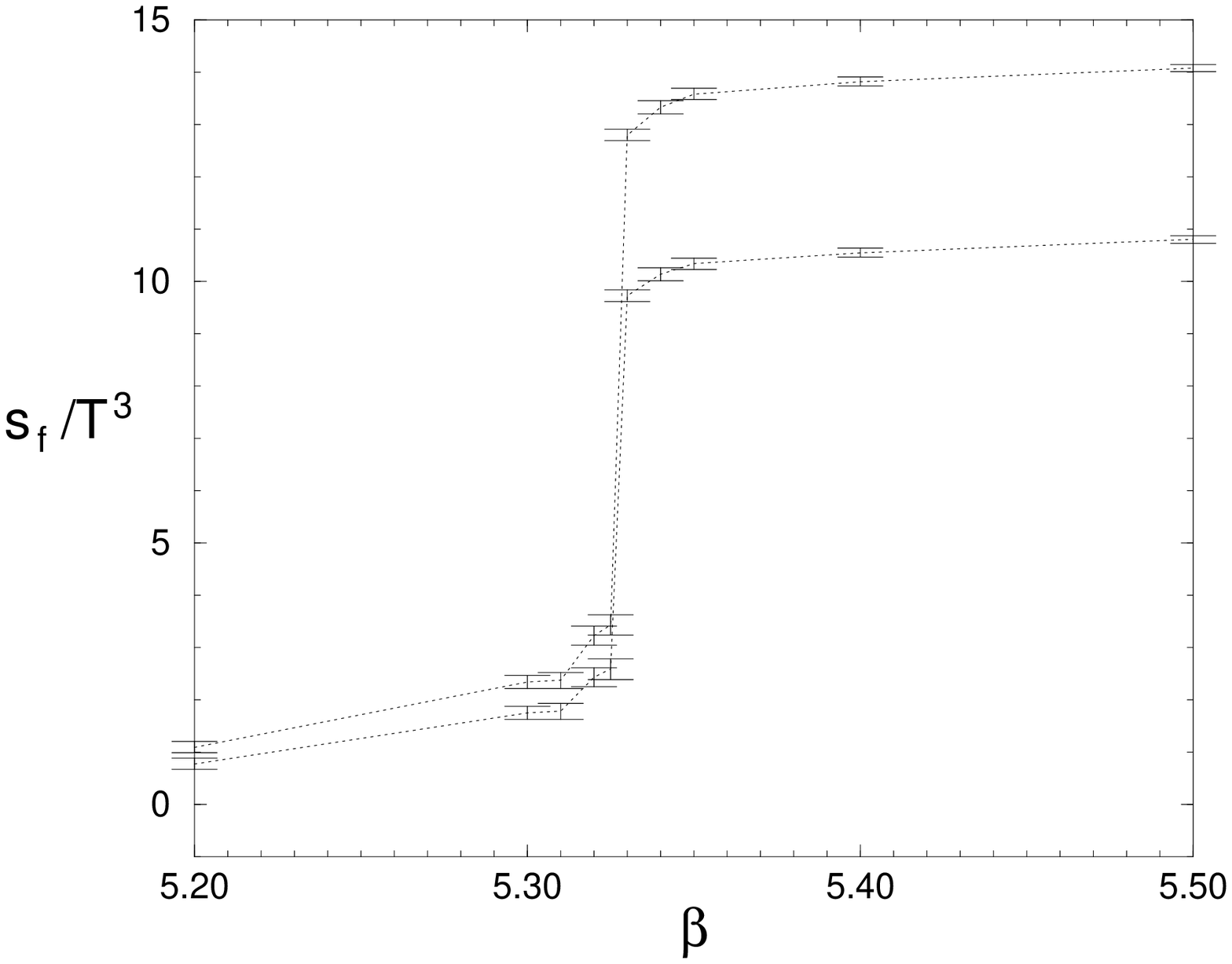}
\caption{Fermionic entropy density using the data from
         \protect\cite{KogutLS}, for two massless flavors on a
         $8^3\times4$ lattice at $\gamma=10$.  The upper line is for
         the tree level results and the lower line the one-loop
         corrected results.}
\label{fig:quark_entropy}
\end{center}
\end{figure}

Even in the absence of such zero temperature measurements the
$8^3\times 4$ lattice data from \cite{KogutLS} can be used with
Eq.~(\ref{xqcd-entropy}) to extract the entropy density.  The results
for the gluonic and fermionic entropy densities are plotted separately
in Figs.~\ref{fig:glue_entropy} and \ref{fig:quark_entropy},
respectively.  The upper lines are the tree level results and the
lower lines are the one-loop corrected results.  The figure shows a
clear first order transition consistent with the measurements of the
Wilson line and the chiral order parameter in the same study.
Theoretically \cite{Pisarski-Wilczek}, one expects a second order
transitions in the continuum theory.  The first order transition here
is most likely due to the small size of the lattice used.  Similar
results are found in conventional lattice QCD with staggered quarks on
lattices of that size.  Clearly, larger lattices will be necessary to
determine the nature of the chiral phase transition and the equation
of state of the quark-gluon plasma in the continuum limit.

\section{Mean Field Approximation}
\label{sec:MFA}
In this section, the mean field treatment of the four-fermion
interactions is discussed in greater detail and then used to
explicitly calculate the regularization dependence of the four-fermion
coupling $G=1/\gamma$ to lowest order in the gauge coupling $g$.  For
these purposes it is best to work with the model where the
$\sigma\bar\psi\psi$ vertex and the dynamical fermion mass are
proportional to $\sqrt{G}$. This is accomplished by rescaling the
auxiliary fields such that the action is
\begin{eqnarray}
S &=& \sum_x\biggl[\beta_{\sigma}\sum_{i<j}P_{ij}(x) 
          + \beta_{\tau}\sum_jP_{0j}\biggr]
+ \sum_{a=1}^{N_f/4}\biggl\{\sum_{x,y}\bar\chi^a(x)[\sum_{j=1}^3
{\cal M}_j(x,y) + \gamma_F{\cal M}_0(x,y)]\chi^a(y)\nonumber\\ 
& \ &\quad {}+
\sum_x\bar\chi^a(x){\sqrt{G}\over16}\!\!\sum_{<x,\tilde x>}
[\sigma(\tilde x)+i\varepsilon(x)\pi(\tilde x)]\chi^a(x) \biggr\} +
{N_f\over8}\sum_{\tilde x}[\sigma^2(\tilde x) + \pi^2(\tilde x)].
\end{eqnarray}
Using identities such as $\sqrt{G}{\textstyle{N_f\over4}}
\langle\bar\chi(\sigma+i\varepsilon\pi)\chi\rangle =2(1
-{\textstyle{N_f\over8}}\langle\sigma^2+\pi^2\rangle)$ it is easy to
verify that the final expressions of the previous section,
(\ref{xqcd-energy}) -- (\ref{xqcd-entropy}), will remain unchanged.
  
In a mean field (or a large $N_f$) approximation of the four-fermion
interactions the auxiliary fields take on the values
\begin{eqnarray}
\langle\sigma\rangle &=& \sqrt{G}\langle\psi^a\bar\psi^a\rangle \\
                     &=& \sqrt{G}\mbox{Tr}[S_F'(x-x)], \\ 
\langle\pi\rangle    &=& \sqrt{G}\langle\psi^a\gamma_5\bar\psi^a\rangle \\ 
                     &=& \sqrt{G}\mbox{Tr}[\gamma_5S_F'(x-x)],
\end{eqnarray}
where $S_F'$ is the full quark propagator (including the gluonic
contributions):
\begin{equation}
\label{fullpropogator}
{S_F'}^{-1}(p)=i\sum_{\lambda}\gamma_{\lambda}
\sin(p_{\lambda}a_{\lambda})/a_{\lambda} + \langle\sigma\rangle
+ i\gamma_5\tau_3\langle\pi\rangle + \Sigma(p) + O(a).
\end{equation}
Here $\Sigma(p)$ is the quark self-energy of 
Sec.~\ref{sec:selfenergy} evaluated at $m^2=\langle\sigma\rangle^2+
\langle\pi\rangle^2$.  The lowest order gluonic
contributions to $\langle\sigma\rangle$ are shown in
Fig.~\ref{fig:fyngap}.
\begin{figure}
\begin{center}
\begin{eqnarray}
\langle\sigma\rangle \quad =
\qquad\parbox{30mm}{
\begin{fmfgraph}(60,40)
\fmfpen{thin}
\fmfbottom{i}
\fmftop{o}
\fmf{phantom}{o,v4}
\fmf{dashes,tension=.75}{i,v}
\fmf{plain,tension=0.25,left}{v,v4,v}
\end{fmfgraph}
} + \qquad\parbox{30mm}{
\begin{fmfgraph}(60,40)
\fmfpen{thin}
\fmfbottom{i}
\fmftop{o}
\fmfleft{l}
\fmfright{r}
\fmf{dashes,tension=.75}{i,v1}
\fmf{phantom}{o,v4}
\fmf{plain,tension=0.5,right=0.4}{v2,v1,v3}
\fmf{plain,tension=0.5,left=0.4}{v2,v4,v3}
\fmf{phantom,tension=0.45}{l,v2}
\fmf{phantom,tension=0.45}{v3,r}
\fmf{photon,tension=0.01}{v2,v3}
\fmfdot{v2,v3}
\end{fmfgraph}
} + \qquad\parbox{30mm}{
\begin{fmfgraph}(60,40)
\fmfpen{thin}
\fmfbottom{i}
\fmftop{o}
\fmf{dashes}{i,v}
\fmf{plain,tension=0.3,left}{v,v1,v}
\fmf{photon,tension=0.5,left}{v1,o,v1}
\fmfdot{v1}
\end{fmfgraph}
} \nonumber
\end{eqnarray}
\caption{Contributions to $\langle\sigma\rangle$ up through $O(g^2)$.}
\label{fig:fyngap}
\end{center}
\end{figure}

Now the idea of the $\chi$QCD model is to choose the value of $G$ to
be very small, so that the breaking of the chiral symmetry is due to
the {\it long distance} behavior of the gauge fields (as opposed to the
spontaneous symmetry breaking in Gross-Neveu\cite{GrossN74} and
Nambu--Jona-Lasinio\cite{NambuJL61} type models).  Since this is
inherently a non-perturbative effect it will not be found in
perturbation theory at any finite order of the gauge
coupling. Therefore, we will have to rely on the values of
$\langle\sigma\rangle$ and $\langle\pi\rangle$ being determined by
Monte Carlo simulation.

For simplicity of the following discussion choose
$\langle\pi\rangle=0$.  A non-vanishing value of
$\langle\sigma\rangle$ then signals dynamical chiral symmetry
breaking.  Following the one-loop calculation of the quark propagator
in Sec.~\ref{sec:selfenergy}, the dynamically generated quark mass
(determined by the pole in $S'_F(p)$) is then
\begin{eqnarray}
\sqrt{G}\langle\sigma\rangle\biggl[1 + g^2{4\over3}\Sigma_2(\xi) 
                                     + O(g^4)\biggr].
\end{eqnarray}
Hence, the regularization dependence of this physical mass can be absorbed
into the bare four-fermion coupling $G$ by taking 
\begin{eqnarray}
\label{mascaling}
\sqrt{G(\xi)}\langle\sigma\rangle 
       = \sqrt{G}\langle\sigma\rangle\biggl[1 + g^2 C_m(\xi) + O(g^4)\biggr],
\end{eqnarray}
with $C_m(\xi)$ evaluated at $m=\sqrt{G}\langle\sigma\rangle$.  

The $\xi$ dependence of $G$ can also be determined by calculating the
one-loop gluonic contributions to the $\sigma\bar\psi\psi$ vertex
shown in Fig.~\ref{fig:sigffvertex}.
\begin{figure}
\begin{center}
\begin{eqnarray}
\Gamma_{\sigma\bar\psi\psi} \quad =
\qquad\parbox{30mm}{
\begin{fmfgraph}(60,40)
\fmfpen{thin}
\fmfleft{i}
\fmfright{o1,o2}
\fmf{dashes}{i,v}
\fmf{fermion}{o1,v}
\fmf{fermion}{v,o2}
\end{fmfgraph}
} + \qquad\parbox{30mm}{
\begin{fmfgraph}(60,40)
\fmfpen{thin}
\fmfleft{i}
\fmfright{o1,o2}
\fmf{dashes}{i,v1}
\fmf{fermion,tension=0.6}{v2,v1}
\fmf{fermion,tension=0.6}{v1,v3}
\fmf{photon,tension=0}{v2,v3}
\fmf{plain}{v2,o1}
\fmf{plain}{v3,o2}
\fmfdot{v2,v3}
\end{fmfgraph}
} \nonumber
\end{eqnarray}
\caption{Contributions to the $\sigma\bar\psi\psi$ vertex up through $O(g^2)$.}
\label{fig:sigffvertex}
\end{center}
\end{figure}
It suffices to consider external momenta less than $1/a$.  It is then
easy to see that the one loop gluonic correction to
$\Gamma_{\sigma\bar\psi\psi}$ is identical to the scalar part of the
one loop fermion self energy calculated in 
Sec.~\ref{sec:selfenergy}:
\begin{eqnarray}
\Gamma_{\sigma\bar\psi\psi} &=& {\sqrt{G}\over N_f}\biggl[ 1 - 
    g^2{N^2-1\over 2N}\sum_{\mu}\int_q 
    \gamma_{\mu}S_F(p+q)S_F(p+k+q)\gamma_{\mu}D(q) \nonumber \\
& \ & \qquad\qquad\qquad\qquad\qquad\qquad\qquad
      \times\cos((p_{\mu}+q_{\mu}/2)a_{\mu})
      \cos((p_{\mu}+k_{\mu}+q_{\mu}/2)a_{\mu})\biggr]\\
          &=& {\sqrt{G}\over N_f}\biggl[1 + g^2{4\over3}
          \int_q{4-{\textstyle{1\over2}}\Delta_1(1)\over
          2\Delta_1(\xi)(\Delta_2(\xi)+G\langle\sigma\rangle^2)/a^4} 
             + \mbox{terms higher O(pa, ka)}\biggr] \\
          &=&  {\sqrt{G}\over N_f}\biggl[1 +
    g^2{4\over3}\Sigma_2(\xi)\biggr], 
\end{eqnarray}
where $\Sigma_2$ is the same as for ordinary QCD [Eq.~(\ref{Sigma2})]
with $m=\sqrt{G}\langle\sigma\rangle$.  Defining
$\Gamma_{\sigma\bar\psi\psi}(\xi) =
Z_{\psi}^{-1}(\xi)\Gamma_{\sigma\bar\psi\psi}^{sym}$, the $\xi$
dependence is absorbed by taking $\sqrt{G(\xi)}=\sqrt{G}[1 +
g^2{\textstyle{4\over3}}(\Sigma_{1,\sigma}(\xi)-\Sigma_2(\xi))] =
\sqrt{G}[1 + g^2C_m(\xi)]$, in agreement with Eq.~(\ref{mascaling}).  

Differentiating this with respect to $\xi$ yields
\begin{eqnarray}
{1\over G}\biggl({\partial G\over\partial\xi}\biggr)_{\xi=1} =
  2 g^2 \biggl({\partial C_m\over\partial\xi}\biggr)_{\xi=1}. 
\end{eqnarray}
The scaling of $G$ with the spatial lattice spacing (at
fixed $\xi$) is then given by $G(a)=G[1 + g^2{\textstyle{8\over3}}
(\Sigma_{1,\sigma}-\Sigma_2)_{DIV}]$, where $(\Sigma_{1,\sigma})_{DIV}
= -1/(8\pi^2)\ln(a)$ and $(\Sigma_2)_{DIV} = -1/(2\pi^2)\ln(a)$ are
the divergent parts of $\Sigma(p)$ as $a\to0$ \cite{Lee94}.  This
yields
\begin{eqnarray}
{a\over G}{\partial G\over\partial a} &=& g^2 {2\over2\pi^2} + O(g^4). 
\end{eqnarray}

\section{Lattice QCD Thermodynamics with Nonzero Quark Masses}
\label{sec:pertQCD}
This section briefly reviews the perturbative thermodynamics of
lattice QCD, including the contributions due to nonzero bare quark
masses.  The necessary one-loop calculations of
\cite{Karsch82,Trinchero,Karsch89} are then extended to nonzero quark
mass and the results discussed.

On an anisotropic lattice the action for QCD with $N_f$ (degenerate)
flavors of staggered fermions is
\begin{eqnarray}
S = \sum_x\biggl[\beta_{\sigma}\sum_{i<j}P_{ij}(x) 
  + \beta_{\tau}\sum_jP_{0j}\biggr] 
  + \sum_{a=1}^{N_f/4}\sum_{x,y}a^2a_{\tau}
  \bar\chi^a(x)Q(x,y)\chi^a(y),
\end{eqnarray} 
with
\begin{eqnarray}
 Q(x,y) = \sum_{j=1}^3 
  {\cal M}_j(x,y)+\gamma_F{\cal M}_0(x,y)+\delta_{x,y}m_0a.
\end{eqnarray} 
The separate gauge couplings $\beta_{\sigma}=6/g^2_{\sigma}$,
$\beta_{\tau}=6/g^2_{\tau}$, as well as the extra parameter $\gamma_F$,
are necessary to maintain Euclidean invariance (i.~e.~regularization
independence) in the continuum limit.  The functional dependence of
these parameters on $\xi\equiv a/a_{\tau}$ is fixed by requiring the
theory to be independent of $\xi$ in the continuum limit.

As shown in Sec.~\ref{sec:therm}, to calculate thermodynamic
quantities one needs
\begin{eqnarray}
\label{QCD_dSdxi}
\biggl\langle\xi{\partial S\over\partial\xi}\biggr\rangle &=& 
N_{\sigma}^3N_{\tau}\biggl[
 \xi{\partial\beta_{\sigma}\over\partial\xi}\langle P_{ss}\rangle
 + \xi{\partial \beta_{\tau}\over\partial\xi}\langle P_{st}\rangle \nonumber\\
 &\ & \ {}+ \xi{\partial \gamma_F\over\partial\xi}{N_f\over4}
           \langle\bar\chi{\cal M}_0\chi\rangle
 + \xi{\partial m_0a\over\partial\xi}{N_f\over4}\langle\bar\chi\chi\rangle
 + \biggl(-1 + \xi{\partial Z_{\psi}\over\partial\xi}\biggr)
 {N_f\over4}\langle\bar\chi Q\chi\rangle\biggr],
\end{eqnarray} 
and
\begin{eqnarray}
\label{QCD_dSda}
\biggl\langle a{\partial S\over\partial a}\biggr\rangle &=& 
N_{\sigma}^3N_{\tau}\biggl[
 a{\partial\beta_{\sigma}\over\partial a}\langle P_{ss}\rangle
 + a{\partial \beta_{\tau}\over\partial a}\langle P_{st}\rangle \nonumber\\
 &\ & \ {}+ a{\partial \gamma_F\over\partial a}{N_f\over4}
    \langle\bar\chi{\cal M}_0\chi\rangle
 + a{\partial(m_0a)\over\partial a}{N_f\over4}\langle\bar\chi\chi\rangle
+ \biggl(3 + {\partial Z_{\psi}\over\partial\ln a}\biggr)
 {N_f\over4}\langle\bar\chi Q\chi\rangle
\biggr].
\end{eqnarray} 
The averages appearing above can be evaluated by Monte Carlo
simulation.  $\langle\bar\chi Q\chi\rangle=-\tr(QQ^{-1})=-3$, for 3
colors of quarks.  Hence, when subtracting $T=0$ measurements on
symmetric lattices the last terms appearing in Eqs.~(\ref{QCD_dSdxi})
and (\ref{QCD_dSda}) will not contribute.  In general, the various
derivatives of the coefficients may also need to be calculated
nonperturbatively \cite{KarschES97}.  As a first step, however, they
can be estimated by lattice perturbation theory.

Considering first the $\xi$ dependence:
\begin{eqnarray}
\beta_{\sigma}(\xi) &=& {6\over\xi g^2}
                   \biggl[1 + C_{\sigma}(\xi)g^2 + O(g^4)\biggr], \\ 
\beta_{\tau}(\xi) &=& {6\xi\over g^2}
                 \biggl[1 + C_{\tau}(\xi)g^2 + O(g^4)\biggr], \\ 
\gamma_F(\xi) &=& \xi\biggl[1 + C_F(\xi)g^2 + O(g^4)\biggr], \\ 
m_0(\xi)       &=& m_0 \biggl[ 1 + C_m(\xi)g^2 + O(g^4)\biggr].
\end{eqnarray}
[Here $g \equiv g_{\sigma,\tau}(\xi=1)$ and $m_0 \equiv m_0(\xi=1)$.]
The coefficients $C_{\sigma}$, $C_{\tau}$, $C_F$ and $C_m$ all vanish
at $\xi=1$ where most Monte Carlo simulations are performed.  However,
their derivatives with respect to $\xi$ are nonzero at $\xi=1$.
They are calculated in Secs.~\ref{sec:vacuum} and
\ref{sec:selfenergy}.

The implicit scaling of the gauge coupling $g$ and the bare quark mass
$m_0$ with the spatial lattice spacing is governed by the
renormalization group equations:
\begin{eqnarray}
-a{\partial g\over\partial a} &=& \beta(g) = -\beta_0g^2-\beta_1g^5+O(g^7),\\
\beta_0 &=& {11\over16\pi^2}\biggl(1-{2N_f\over33}\biggl), \qquad
\beta_1 = {1\over16\pi^2}\biggl[102-\biggl(10+{8\over3}\biggr)N_f\biggl], \\
a{\partial m_0\over\partial a} &=& 
                  m_0 \gamma(g) = m_0\biggl[\gamma_0g^2 + O(g^4)\biggr], \\
\gamma_0 &=& {1\over 2\pi^2}.
\end{eqnarray}
$\beta(g)$ is universal (regularization independent) up through
$O(g^5)$, and $\gamma(g)$ is universal through $O(g^2)$.
At $\xi=1$
\begin{eqnarray}
 a{\partial\beta_{\sigma}\over\partial a}
 = a{\partial \beta_{\tau}\over\partial a}
 = a{\partial (6/g^{2})\over\partial a} 
 = -12\beta_0 + O(g^2),
\end{eqnarray}
and 
\begin{eqnarray}
a{\partial\gamma_F\over\partial a} = 0.
\end{eqnarray}

The final perturbative expressions for the energy, pressure, and entropy (at
$\xi=1$) are:
\begin{eqnarray}
\label{finalqcdenergy}
\epsilon &=& T^4 N_{\tau}^4
  \biggl\{ {6\over g^2}\biggl(1 - g^2{\partial C_{\sigma}
                     \over\partial\xi}\biggr)
  \langle P_{ss}\rangle + {6\over g^2}\biggl(-1 - g^2{\partial C_{\tau}
        \over\partial\xi}\biggr)\langle P_{st}\rangle \nonumber \\
 & \ & \quad  {}+  \biggl(1 + g^2{\partial
    C_F\over\partial\xi}\biggr){N_f\over4}
           \langle\tr{\cal M}_0 Q^{-1}\rangle \nonumber \\   
 & \ & \quad   {}+ m_0ag^2{\partial C_m\over\partial\xi}{N_f\over4}
              \langle\tr Q^{-1}\rangle
 + 3{N_f\over4}\biggl(-1 + {\partial Z_{\psi}\over\partial\xi}\biggr) 
 \biggr\},  \\
\label{finalqcdpressure}
p &=& {\epsilon\over3} + {T^4\over3} N_{\tau}^4
     \biggl\{ 12\beta_0\biggl(\langle P_{ss}\rangle 
     + \langle P_{st}\rangle\biggr) \nonumber \\
 & \ &\quad {}+ m_0a(1 + g^2\gamma_0) {N_f\over4}\langle\tr Q^{-1}\rangle
 + 3{N_f\over4}\biggl(3 + 
 {\partial Z_{\psi}\over\partial\ln a}\biggr)\biggr\}, \\ 
s &=& {4T^3\over3} N_{\tau}^4
  \biggl\{ {6\over g^2}\biggl(1 - g^2{\partial C_{\sigma}
  \over\partial\xi}\biggr) \langle P_{ss}\rangle +
  {6\over g^2}\biggl(-1 - g^2{\partial C_{\tau}
  \over\partial\xi}\biggr)\langle P_{st}\rangle 
  + 3\beta_0\biggl(\langle P_{ss}\rangle 
  + \langle P_{st}\rangle\biggr) \nonumber \\
 & \ & \quad {}+ \biggl(1 + g^2{\partial
  C_F\over\partial\xi}\biggr){N_f\over4}
  \langle\tr{\cal M}_0 Q^{-1}\rangle 
  -{3\over4}{N_f\over4}\biggl(1 - g^2 4{\partial
  Z_{\psi}\over\partial\xi}-g^2{\partial Z_{\psi}\over\partial\ln
  a}\biggr) \nonumber \\   
 & \ & \quad   {}+ m_0a
  \biggl(1 + g^2\gamma_0+g^24{\partial
  C_m\over\partial\xi}\biggr) {N_f\over 16}\langle\tr Q^{-1}\rangle 
\biggr\} \\
 &=& {4T^3\over3} N_{\tau}^4
  \biggl\{ {6\over g^2}\biggl[1 + g^2{1\over2}\biggl({\partial C_{\tau}
  \over\partial\xi}-{\partial C_{\sigma}\over\partial\xi}\biggr)
  \biggr] \biggl[\langle P_{ss}\rangle -\langle P_{st}\rangle\biggr] 
  \nonumber \\
 & \ & \quad {}+ \biggl(1 + g^2{\partial
  C_F\over\partial\xi}\biggr){N_f\over4}
  \biggl[\langle\tr{\cal M}_0 Q^{-1}\rangle -{3\over4}+ 
  {1\over4}m_0a\langle\tr Q^{-1}\rangle \biggr] \biggr\}.
\label{finalqcdentropy}
\end{eqnarray}
The $m_0\approx 0$ relations given by Eqs.~(\ref{gauge_sum_rule}),
(\ref{self-energy_sum_rule}) and (\ref{mass_sum_rule}) have been used to
simplify the expression for the entropy density.  In most cases one is
interested in measuring the energy density and pressure relative to
the $T=0$ vacuum, by subtracting measurements on symmetric lattices.
In which case, the last terms in Eqs.~(\ref{finalqcdenergy}) and
(\ref{finalqcdpressure}) will drop out.
Using
$\langle\tr Q Q^{-1}\rangle =3$ it follows that $\langle\tr{\cal M}_0
Q^{-1}\rangle_{sym} = {\textstyle{3\over4}}
-{\textstyle{1\over4}}m_0a\langle\tr Q^{-1}\rangle_{sym}$.  Hence, it
is easy to verify that the expression for the entropy vanishes on an
isotropic, symmetric lattice.

The one loop perturbative calculations of the various coefficients
appearing above, and in particular their dependence on the bare quark
mass $m_0$, are presented in the following two sections.

\subsection{Vacuum Polarization}
\label{sec:vacuum}

Define $C_{\sigma}(\xi) \equiv C_{\sigma}^G(\xi) + C_{\sigma}^F(\xi)$ 
and $C_{\tau}(\xi) \equiv C_{\tau}^G(\xi) + C_{\tau}^F(\xi)$
where the superscripts G and F refer to the gauge and fermionic 
contributions to the QCD vacuum polarization, respectively. 
The $\xi$ dependence is most easily calculated using a background field 
method\cite{DashenGross,Karsch82}; 
i.~e.~by expanding the gauge action around a classical background gauge
configuration and requiring the effective action to be independent of $\xi$.   

The pure gauge contributions have been calculated by Karsch\cite{Karsch82}.
They yield
$({\partial C_{\sigma}^G/\partial\xi})_{\xi=1} = 0.20161$ and
$({\partial C_{\tau}^G/\partial\xi})_{\xi=1} = -0.13195$.
The fermionic contributions are given solely by the one-loop vacuum
polarization graphs of Eqs.~(\ref{vac_a}) and (\ref{vac_b}).
These have been evaluated  previously by Trinchero for
massless quarks\cite{Trinchero}.  Following his notation,
\begin{eqnarray}
\Pi_{ij} &=& (-k_i k_j I_{\sigma} + \delta_{ij})
({\bf k^2}I_{\sigma}+k^2_0I_{\tau}), \\
\Pi_{\mu0} &=& (-k_{\mu} k_0 + \delta_{\mu0} k^2)I_{\tau}, 
\end{eqnarray}
with
\begin{eqnarray}
I_{\sigma}(\xi) &=&  -{N_f\over2}\int_{q/2}
{\cos^2(q_1a)\cos^2(q_2a)-{1\over3}\cos(2q_1a)\cos(2q_2a) \over
[\Delta_2(\xi) + m_0^2a^2]^2/a^4},      \\
I_{\tau}(\xi) &=& -{N_f\over2}\int_{q/2}{\cos^2(q_1a)\cos^2(q_0a_{\tau})
-{1\over3}\cos(2q_1a)\cos(2q_0a_{\tau})\over [\Delta_2(\xi) 
+ m_0^2a^2]^2/a^4}. 
\end{eqnarray}
$C_{\sigma}^F$ and  $C_{\tau}^F$ are related to these integrals by
\begin{eqnarray}
C_{\sigma}^F = 2[I_{\sigma}(\xi)-I_{\sigma}(1)], \qquad\qquad
C_{\tau}^F = 2[I_{\tau}(\xi)-I_{\tau}(1)].
\end{eqnarray}

\begin{figure}
\begin{center}
\leavevmode
\epsfxsize=4.2in
\epsfbox{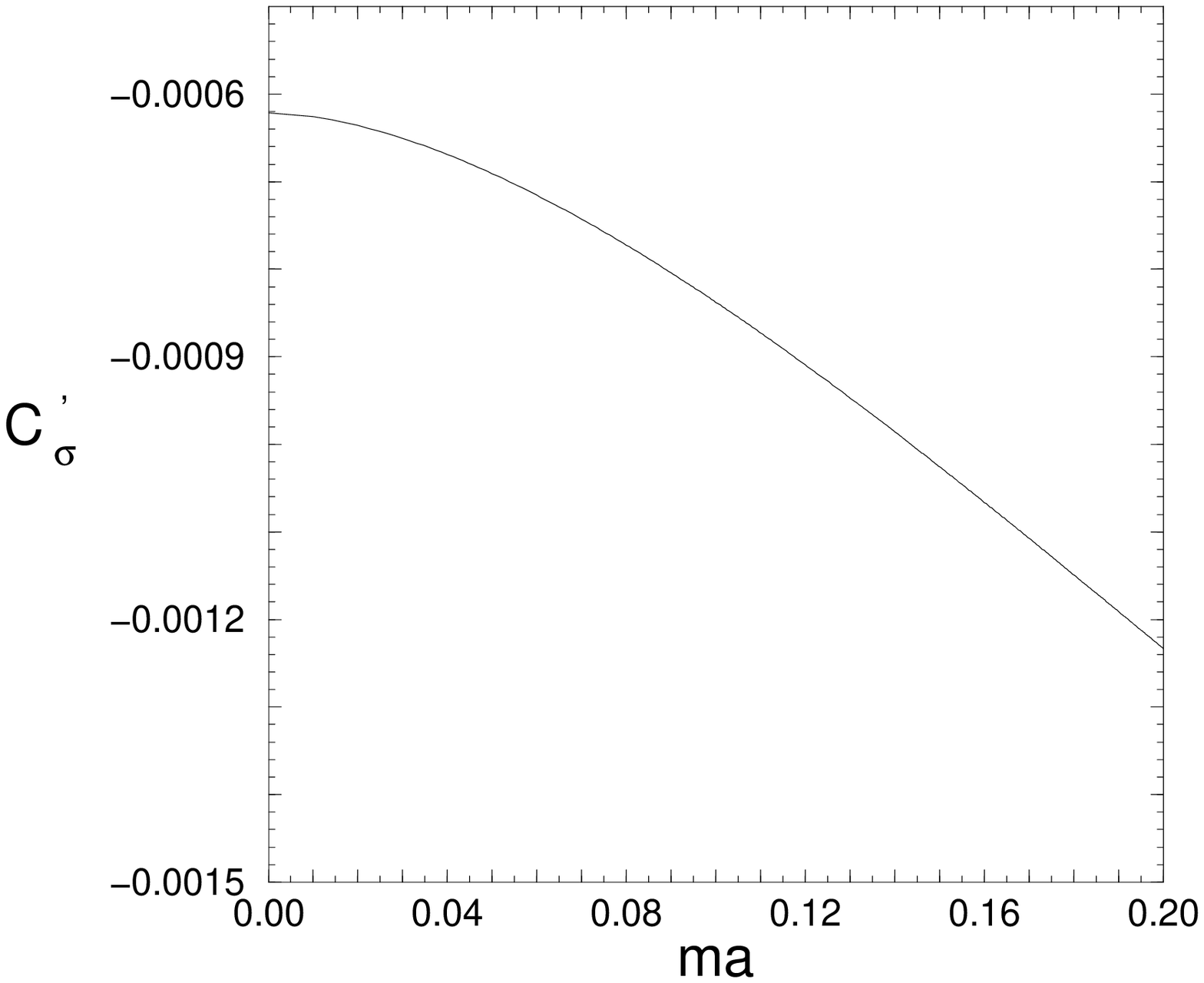}
\caption{$(\partial C_{\sigma}^F/\partial\xi)_{\xi=1}$ as a function
         of the mass (for two quark flavors).}
\label{fig:csigf}
\leavevmode
\epsfxsize=4.2in
\epsfbox{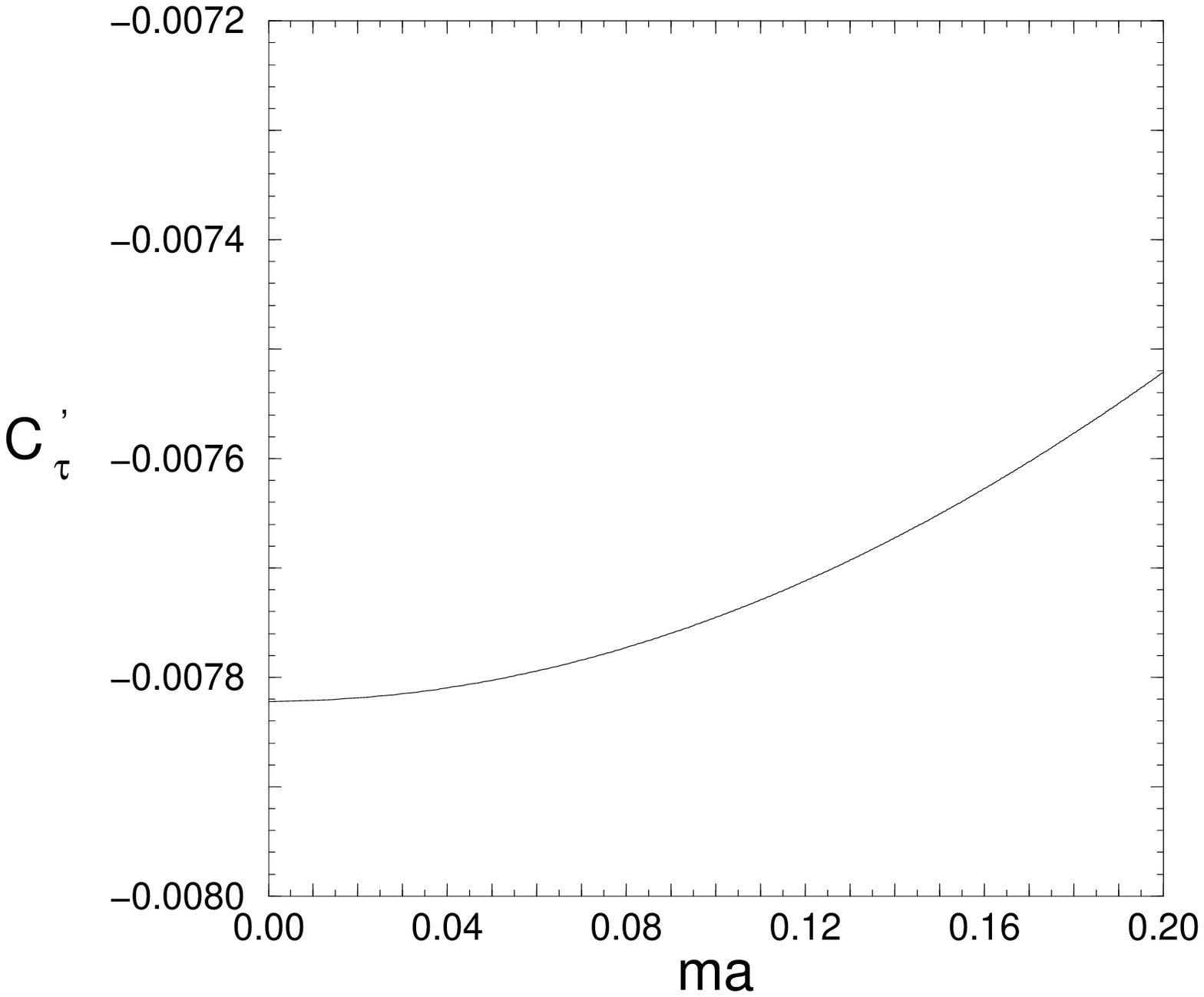}
\caption{$(\partial C_{\tau}^F/\partial\xi)_{\xi=1}$ as a function
         of the mass (for two quark flavors).}
\label{fig:ctauf}
\end{center}
\end{figure}

For $m_0=0$ I find
$({\partial C_{\sigma}^F/ \partial\xi})_{\xi=1} 
= -{\textstyle{N_f\over 2}}0.000622$ and 
$({\partial C_{\tau}^F/ \partial\xi})_{\xi=1} 
= -{\textstyle{N_f\over 2}}0.007822$,
in agreement with \cite{Trinchero}.
The behaviors of $(\partial C_{\sigma}^F/\partial\xi)_{\xi=1}$ and
$(\partial C_{\tau}^F/\partial\xi)_{\xi=1}$ (for $N_f=2$) as a
function of $m_0a$ are shown in Figs.~\ref{fig:csigf} and
\ref{fig:ctauf}, respectively.
As $m_0a$ increases from zero to 0.1, $\partial C_{\sigma}/\partial\xi
= \partial C_{\sigma}^G/\partial\xi+
\partial C_{\sigma}^F/\partial\xi$
decreases 0.1\% and $\partial C_{\tau}/\partial\xi = \partial
C_{\tau}^G/\partial\xi+\partial C_{\tau}^F/\partial\xi$ increases
0.05\% (per $N_f/2$ flavors).  Therefore, in most cases when $m_0a$ is
small these deviations can be ignored and the massless quark values
used.

The invariance of the string tension\cite{Karsch82} as well as the
vanishing of the gluonic entropy at zero temperature requires 
\begin{eqnarray}
\label{gauge_sum_rule}
\biggl({\partial \beta_{\sigma}\over\partial\xi}\biggr)_{\xi=1} +
\biggl({\partial \beta_{\tau}\over\partial\xi}\biggr)_{\xi=1} =
-{1\over2}{\partial (6/g^{2})\over\partial \ln a}.
\end{eqnarray}
To lowest order in $g$ this reads $(\partial C_{\sigma}/\partial\xi)_{\xi=1} +
(\partial C_{\tau}/\partial\xi)_{\xi=1} = \beta_0$.  This relation no longer
holds at nonzero $m_0a$.  However, given the relatively small
variations in $(\partial C_{\sigma}/\partial\xi)_{\xi=1}$ and $(\partial
C_{\tau}/\partial\xi)_{\xi=1}$ it is still effectively satisfied for
small bare masses.

\subsection{Quark Propagator}           
\label{sec:selfenergy}

$\gamma_F(\xi)$ can be found by calculating the one loop corrections
to the fermion propagator (i.e.~the fermionic self-energy) and then
demanding rotational invariance of the propagator in the continuum
limit.  This has also been carried out by Karsch for massless
fermions\cite{Karsch89}, and will be extended here to the more general
case.

\begin{figure}
\begin{center}
\leavevmode
\epsfxsize=4in
\epsfbox{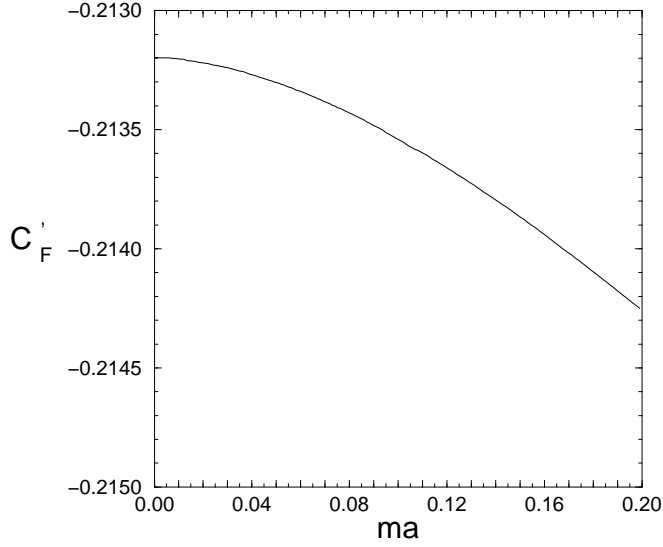}
\caption{$(\partial C_F/\partial\xi)_{\xi=1}$ as a function of the mass.}
\label{fig:gammaF}
\end{center}
\end{figure}

Consider
\begin{eqnarray}
\Gamma^{(2)}(p) \equiv S_F^{-1}(p) + \Sigma(p).
\end{eqnarray}
To lowest order in $g$, $-\Sigma(p)$ is given by the sum of the two terms:
\begin{eqnarray}
\parbox{20mm}{\begin{fmfgraph}(50,50)
\fmfpen{thin}
\fmfleft{i}
\fmfright{o}
\fmf{plain}{i,v1}
\fmf{fermion,tension=.5}{v1,v2}
\fmf{plain}{v2,o}
\fmf{photon,tension=0,left}{v1,v2}
\fmfdot{v1,v2}
\end{fmfgraph}}    
&=& - g^2{N^2-1 \over 2N}\sum_{\mu}\int_q \gamma_{\mu}S_F(p+q)\gamma_{\mu}
D(q)\cos^2(p_{\mu}a_{\mu}+q_{\mu}a_{\mu}/2), \\
\parbox{20mm}{\begin{fmfgraph}(50,50)
\fmfpen{thin}
\fmfleft{i}
\fmfright{o}
\fmf{fermion}{i,v,o}
\fmf{photon}{v,v}
\fmfdot{v}
\end{fmfgraph}}
&=& g^2{N^2-1 \over 2N}
{1\over2}\sum_{\mu}a_{\mu}i\gamma_{\mu}\sin(p_{\mu}a_{\mu})\int_q D(q).
\end{eqnarray}
Expanding the self energy in powers of $p_{\nu}$,
\begin{eqnarray}
\Sigma(p) &=& g^2{N^2-1 \over2N}
\biggl[\sum_{j}i\gamma_{j}p_{j}\Sigma_{1,\sigma} 
+ i\gamma_{0}p_{0}\Sigma_{1,\tau}
+ m_0\Sigma_2\biggr] + \mbox{terms of higher O(pa)},
\end{eqnarray}
with
\begin{eqnarray}
\Sigma_{1,\sigma} &=& \int_q \biggl\{
{ a^4\sin^2(q_1a) \over 2 \Delta_1(\xi)(\Delta_2(\xi)+m_0^2a^2)}
\biggl[ {1\over2} +  { 3 - \cos(q_1a) - 
{\textstyle{1\over2}} \Delta_1(1)
\over \Delta_1(\xi) } \biggr] -  {a^4\over 4\Delta_1(\xi)} \biggr\},
\\
\Sigma_{1,\tau} &=& \int_q \biggl\{
{a^4\sin^2(q_0a_{\tau}) \over 2 \Delta_1(\xi)(\Delta_2(\xi)+m_0^2a^2)}
\biggl[ {1\over2} +  \xi^2{ 3 - \cos(q_0a_{\tau}) - 
{\textstyle{1\over2}} \Delta_1(1)
\over \Delta_1(\xi) } \biggr] -  {a^4\over 4\xi^2\Delta_1(\xi)} \biggr\},
\end{eqnarray}
and
\begin{eqnarray}
\label{Sigma2}
\Sigma_2 = \int_q {4-{\textstyle{1\over2}}\Delta_1(1) 
\over 2\Delta_1(\xi)(\Delta_2(\xi)+ m_0^2a^2)/a^4}.
\end{eqnarray}
Hence,
\begin{eqnarray}
\Gamma^{(2)} &=& \sum_j i\gamma_j p_j 
  \biggl[1 + g^2{N^2-1 \over2N}\Sigma_{1,\sigma}(\xi)\biggr]
  + i\gamma_0 p_0\biggl[ \gamma_F/\xi + 
  g^2{N^2-1 \over2N}\Sigma_{1,\tau}(\xi)\biggr] \nonumber \\
& \ & \quad {}+ m_0\biggr[1 + g^2{N^2-1 \over2N}\Sigma_2(\xi)\biggr]. 
\end{eqnarray}
Demanding rotational invariance of $\Gamma^{(2)}$ in the continuum limit
yields 
\begin{eqnarray}
\gamma_F(\xi) = \xi[1 + g^2 C_F(\xi)+ O(g^4)], 
\end{eqnarray}
with
\begin{eqnarray}
C_F(\xi) 
    = {N^2-1 \over2N}\biggl(\Sigma_{1,\sigma}(\xi)-\Sigma_{1,\tau}(\xi)\biggr).
\end{eqnarray}
For $m_0=0$ I find $({\partial C_F/ \partial\xi})_{\xi=1} = -0.2132$,
in agreement with \cite{Karsch89}.  For nonzero quark masses
$(\partial C_F/\partial\xi)_{\xi=1}$ is shown in
Fig.~\ref{fig:gammaF}.
It decreases less than 0.2\% as $m_0a$ varies from zero to 0.1.
Thus, in most cases the $m_0=0$ value can be safely used.

The remaining $\xi$ dependence of $\Gamma^{(2)}$ can be absorbed by
a wavefunction rescaling:
\begin{eqnarray}
\Gamma^{(2)}(\xi) &=& Z_{\psi}^{-1}(\xi)\Gamma^{(2)}_{sym}, \\
Z_{\psi}(\xi) &=& 1 - g^2{N^2-1 \over2N}\Sigma_{1,\sigma}(\xi),
\end{eqnarray}
and a functional bare mass dependence
\begin{eqnarray}
m_0(\xi) &=& m_0 \biggl[ 1 + g^2 C_m(\xi) + O(g^4) \biggr], \\
C_m(\xi) &=& 
   {N^2-1 \over2N}\biggl(\Sigma_{1,\sigma}(\xi)
- \Sigma_2(\xi) 
\biggr).
\end{eqnarray}
At $m_0=0$ $(\partial\Sigma_{1,\sigma}/\partial\xi)_{\xi=1} =
-0.0368$ and decreases just 0.2\% as $m_0$ increases to 0.1.
The divergent part of $\Sigma_{1,\sigma}$ at $\xi=1$ is
$[\Sigma_{1,\sigma}(1)]_{DIV} = -1/(8\pi^2)\ln(a)$ \cite{Lee94}.
Thus, $\partial Z_{\psi}/\partial\ln a = g^2/(6\pi^2)$.  

\begin{figure}
\begin{center}
\leavevmode
\epsfxsize=4in
\epsfbox{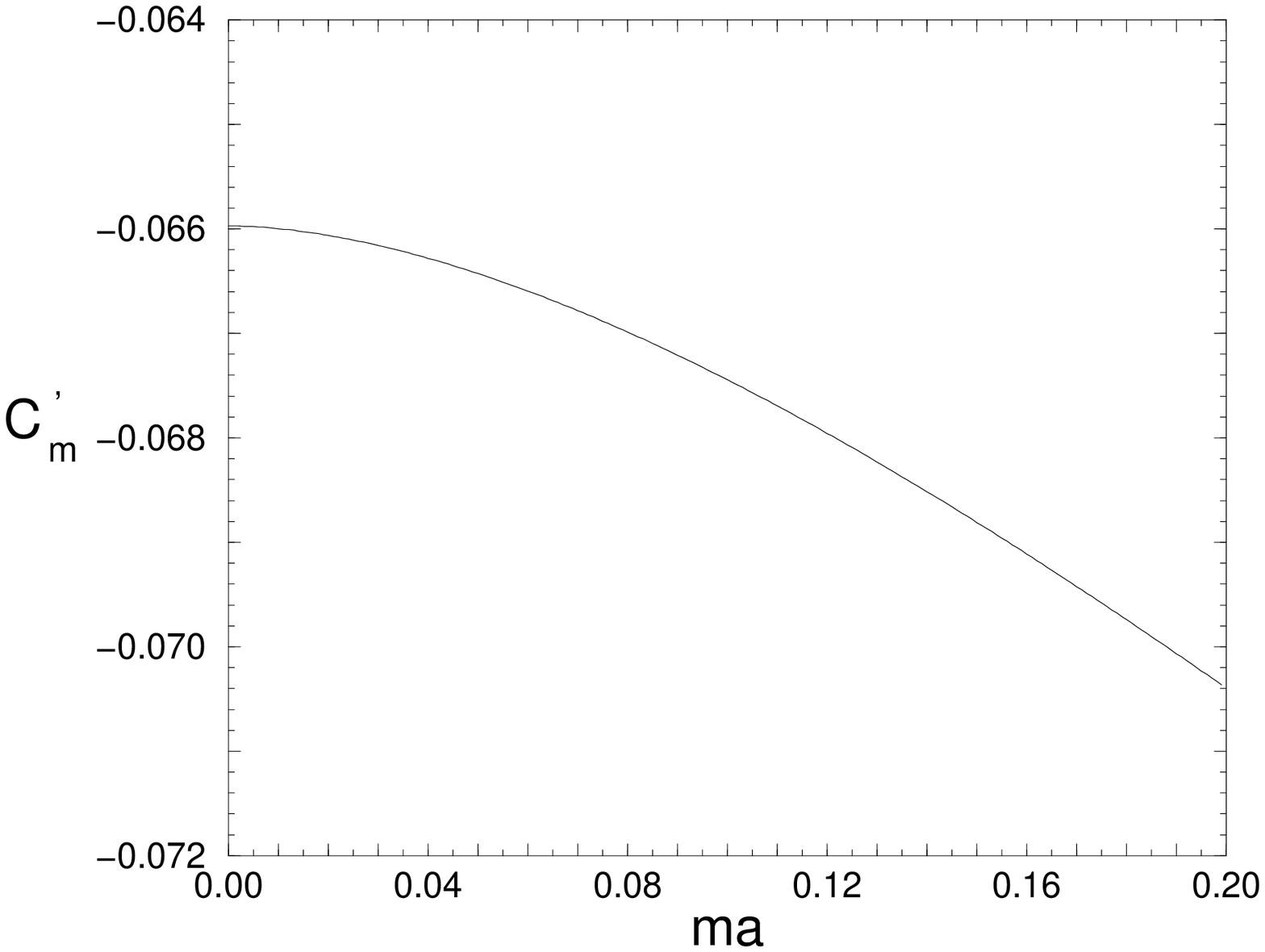}
\caption{$(\partial C_m/\partial\xi)_{\xi=1}$ as a function
of the mass.}
\label{fig:dcmdq}
\end{center}
\end{figure}

Turning to the mass,
\begin{eqnarray}
\biggl({\partial m_0 \over \partial\xi}\biggr)_{\xi=1}
 = m_0 g^2 \biggl({\partial C_m\over\partial\xi}\biggr)_{\xi=1}.
\end{eqnarray}
$(\partial C_m/\partial\xi)_{\xi=1}=-0.066$ at $m_0a=0$. 
Figure \ref{fig:dcmdq} shows the behavior of $(\partial
C_m/\partial\xi)_{\xi=1}$ as a function of $m_0a$.
Its value decreases about 2\% as $m_0a$ increases from zero to 0.1.
This is significantly more variation than seen in $\partial
C_F/\partial\xi$, $\partial C_{\sigma}/\partial\xi$ and $\partial
C_{\tau}/\partial\xi$, and is due mainly to the variation in
$\partial\Sigma_2/\partial\xi$.  
Values of $(\partial C_m/\partial\xi)_{\xi=1}$ are also given in Table
\ref{table:dcmdq} at selected values of $m_0a$.

Requiring the entropy to vanish on an isotropic, symmetric lattice yields
\begin{eqnarray}
\label{self-energy_sum_rule}
g^2\biggl({\partial C_F\over\partial\xi}\biggr)_{\xi=1} +
4\biggl({\partial Z_{\psi}\over\partial\xi}\biggr)_{\xi=1} + {\partial
Z_{\psi}\over\partial\ln a} &=& 0 
\end{eqnarray}
and
\begin{eqnarray}
-g^2\biggl({\partial C_F\over\partial\xi}\biggr)_{\xi=1} +
g^24\biggl({\partial C_m\over\partial\xi}\biggr)_{\xi=1} +{\partial
\ln m_0\over\partial\ln a}  &=& 0,
\label{mass_sum_rule}
\end{eqnarray}
which are satisfied by the numerical values given for
$m_0a\approx 0$.  Like the corresponding sum rule for the gauge
couplings (\ref{gauge_sum_rule}) they are slightly violated at larger
$m_0a$.

\section*{Acknowledgments}
The author would like to thank J. B. Kogut, F. Karsch, and
D. K. Sinclair for helpful discussions.  This work is supported in
part by funds provided by the U.S. Department of Energy (D.O.E.) under
cooperative research agreement DE-FC02-94ER40818.

\section*{Appendix}
On an anisotropic lattice with $\xi\equiv a/a_{\tau}$
the free ($g=0$) lattice propagators are
\begin{eqnarray}
D(k)   &=& {1 \over \sum_{\lambda}
[1-\cos(k_{\lambda}a_{\lambda})]/a^2_{\lambda}}
= {1 \over 2\Delta_1(\xi)/a^2},
\\
S_F(p) &=& {1 \over i\sum_{\lambda}\gamma_{\lambda}
\sin(p_{\lambda}a_{\lambda})/a_{\lambda} + m}
=  {-i\sum_{\lambda}\gamma_{\lambda}
\sin(p_{\lambda})/a_{\lambda} + m \over \Delta_2(\xi)/a^2 + m^2},
\end{eqnarray}
where 
\begin{eqnarray}
\Delta_1(\xi)&\equiv&\sum_{1=1}^3 (1-\cos(k_ia))+\xi^2(1-\cos(k_0a_{\tau})), \\
\Delta_2(\xi)&\equiv&\sum_{1=1}^3 \sin^2(p_ia) + \xi^2\sin^2(p_0a_{\tau}).
\end{eqnarray} 
Integration limits have been denoted by
\begin{eqnarray}
\int_q \equiv \int_{-{\pi\over a}}^{{\pi\over a}}{d^4q\over(2\pi)^4}, 
\qquad\qquad
\int_{q/2} \equiv \int_{-{\pi\over 2a}}^{{\pi\over 2a}}{d^4q\over (2\pi)^4}.
\end{eqnarray}

\begin{table}
\begin{center}
\begin{tabular}{cccc}
$ma$ & $C_m'$& $ma$ & $C_m'$ \\
\hline
  0.000 & -0.0660 &  0.100 & -0.0674  \\
  0.005 & -0.0660 &  0.105 & -0.0676  \\
  0.010 & -0.0660 &  0.110 & -0.0677  \\ 
  0.015 & -0.0660 &  0.115 & -0.0678  \\
  0.020 & -0.0661 &  0.120 & -0.0680  \\
  0.025 & -0.0661 &  0.125 & -0.0681  \\
  0.030 & -0.0662 &  0.130 & -0.0682  \\
  0.035 & -0.0662 &  0.135 & -0.0684  \\
  0.040 & -0.0663 &  0.140 & -0.0685  \\
  0.045 & -0.0664 &  0.145 & -0.0687  \\
  0.050 & -0.0664 &  0.150 & -0.0688  \\
  0.055 & -0.0665 &  0.155 & -0.0690  \\
  0.060 & -0.0666 &  0.160 & -0.0691  \\
  0.065 & -0.0667 &  0.165 & -0.0693  \\
  0.070 & -0.0668 &  0.170 & -0.0694  \\
  0.075 & -0.0669 &  0.175 & -0.0696  \\
  0.080 & -0.0670 &  0.180 & -0.0697  \\
  0.085 & -0.0671 &  0.185 & -0.0699  \\
  0.090 & -0.0672 &  0.190 & -0.0701  \\
  0.095 & -0.0673 &  0.195 & -0.0702  \\
\end{tabular}
\caption{$(\partial C_m/\partial\xi)_{\xi=1}$ at selected values
         of $ma$. }
\label{table:dcmdq}
\end{center}
\end{table}

\end{fmffile}
\end{document}